\documentclass[%
 reprint,
 superscriptaddress,
 amsmath,amssymb,
 aps,
]{revtex4-1}

\usepackage{graphicx}
\usepackage{dcolumn}
\usepackage{bm}

\begin{document}

\preprint{APS/123-QED}

\title{Anderson localization on the Falicov-Kimball model with Coulomb disorder}

\author{R. D. B. Carvalho}
\affiliation{Instituto de F\'isica, Universidade Federal do Rio Grande do
Sul, 91501-970 Porto Alegre, Brazil}

\author{G. M. A. Almeida}
\affiliation{Departamento de F\'isica, Universidade Federal de
Sergipe, 49100-000 S\~ao Cristov\~ao, Brazil} 

\affiliation{Institute for Quantum Science and Technology, University of Calgary, Calgary, Alberta, Canada T2N 1N4} 

\author{A. M. C. Souza}
\affiliation{Departamento de F\'isica, Universidade Federal de
Sergipe, 49100-000 S\~ao Cristov\~ao, Brazil} 

\date{\today}


\begin{abstract}
The role of Coulomb disorder is analysed in the Anderson-Falicov-Kimball model. 
Phase diagrams of correlated and disordered electron systems are calculated within dynamical mean-field theory applied to the Bethe lattice, in which metal-insulator transitions led by structural and Coulomb disorders and correlation can be identified.
Metallic, Mott insulator, and Anderson insulator phases, as well as the crossover between them are studied in this perspective.
We show that Coulomb disorder has a relevant role in the phase-transition behavior as the system
is led towards the insulator regime.

\begin{description}
\item[PACS numbers]
71.10.Fd, 71.23.-k, 71.27.+a, 71.30.+h
\end{description}
\end{abstract}

\pacs{Valid PACS appear here}

\maketitle

\section{\label{secI}Introduction}

The metal-insulator transition (MIT) \cite{imada98} of Anderson type, occurring in systems without interactions having randomly distributed
on-site impurities, is related to electron localization in disordered systems \cite{anderson61}.
On the other hand, the Mott-Hubbard MIT is caused by correlations arising from Coulomb interactions 
in a disorder-free system \cite{mott49}.
Both types of MIT have been extensively explored on its own framework and also when Coulomb correlations
and on-site disordered potentials are simultaneously involved \cite{lee85, belitz94, 
byczuk05, byczuk05-2, souza07, gusmao08, maionchi08}.
Then, if the disorder is intense enough, 
the Mott-Hubbard MIT will naturally take Anderson-localization effects into account.

The interplay between disorder and Coulomb correlations in electronic systems of narrow band is of great interest 
since the experimental control of charge concentration through doping and undesired charged impurities in the background
generates Coulomb disorder scattering \cite{orignac06, shklovskii07, hwang09, sarma13, sarma13-2}.
For models where Coulomb interactions
are considered locally, 
such as the ones described by the Hubbard \cite{hubbard63} and the Falikov-Kimball model \cite{falicov69},
Coulomb disorder should also be taken into account when considering
a random medium which gives rise to Anderson localization phenomena.
Hence, in this work we consider the electron-electron coupling strengths to 
have a random distribution across the lattice and
we investigate how Anderson and Coulomb disorder can mutually 
contribute to the MIT, in the framework of the Falicov-Kimball model.
This model was introduced \cite{falicov69} in order to describe 
the MIT in transition-metal compounds and rare-earth materials. 
It basically assumes that 
there are two species of particles (fermions). One is free to hop among
nearest-neighbor sites, the other one is frozen, and they
experience a local Coulomb interaction.
Apart from being the simplest framework for describing
a MIT induced by electronic correlations,
the Falicov-Kimball model has also a much broader appeal which includes, to name a few, 
the study of crystallization \cite{kennedy86},
order-disorder transition in binary alloys \cite{freericks93},
charge transport \cite{freericks00}, 
and itinerant magnetism \cite{macedo01}. 

In the Anderson-Falicov-Kimball model \cite{byczuk05}, a local random potential is included 
thus disturbing the propagation of free fermions.
In order to turn the model more realistic, we will also consider disorder in the Coulomb repulsion, i.e., the
interaction strength between free and frozen fermion species in each site 
is randomly distributed along the lattice.
The model including both sources of disorder is 
solved within the dynamical mean-field theory (DMFT) \cite{metzner89, georges96} formalism.
A great review on the DFMT applied directly on the Falicov-Kimball model as well
as on the model itself can be found in Ref. \cite{freericks03rev}.

By increasing the Coulomb interaction strength, the model 
captures several aspects of the Mott-Hubbard MIT as the local density of states (LDOS) for mobile fermions 
splits into two sub-bands, leading to a correlation gap at the Fermi level.
An open gap (forbidden energy range) in the density of states at the Fermi level characterizes the Mott-Hubbard MIT. 
The most likely value of the LDOS exhibits a 
discontinuity when the system is going through an Anderson transition.
Along these lines, it is expected that both types of MIT can be detected by evaluating the LDOS. 
Although this quantity is not an order parameter associated with a symmetry breaking 
of the phase transition \cite{byczuk05, souza07},
it discriminates between a metallic and an insulator phase.

In a disordered system, an appropriate average of the random quantities of interest must be performed 
in order to describe the LDOS.
However, the underlying probability distribution function is not completely known
in most of the situations, but only certain averages.
Around the Anderson-MIT boundary, for instance, a witness 
for localization can be provided by evaluating the geometric mean.
This last, in contrast with the
arithmetic mean which is noncritical and does not enable the distinction between localized and extended states, 
gives a better approximation for the LDOS average as it vanishes
at the critical disorder value.
However, by using both averages, each in its the appropriate regime, we are allowed
to provide a good description of the phase diagram and most of the relevant 
information regarding the effects of disorder in the MIT can then be accessed. 
%
%

In what follows, Sec. \ref{secII}, the Anderson-Falicov-Kimball model with Coulomb disorder is 
introduced and solved within the DMFT.
In Sec. \ref{secIII}, the phase diagrams are presented and the interplay between
both sources of disorder are discussed. Final remarks are given in Sec. \ref{secIV}.  

\section{\label{secII}Anderson-Falicov-Kimball model with Coulomb disorder}

The Falicov-Kimball model \cite{falicov69} describes two species of spinless fermions: one is free to move and the other is trapped due to its infinite mass. There is a local coulomb interaction between these two particles and the Pauli exclusion principle assures that
no more than one particle (of a given type) is allowed to occupy the same site. 
Here we consider two kinds of disorder, a local random Anderson-like impurity and Coulomb disorder. This allows us to explore the features in the MIT provided by the competition between these two sources of disorder.
The system's Hamiltonian is then expressed by
\begin{eqnarray} \label{eq01}
H=-\sum_{<ij>} t_{ij} c_{i}^{\dagger}c_{j}+\sum_{i}\epsilon_{i}c_{i}^{\dagger}c_{i}+ \nonumber \\
+\sum_{i} U_{i} f_{i}^{\dagger}f_{i}c_{i}^{\dagger}c_{i}-\mu \sum_{i}c_{i}^{\dagger}c_{i},
\end{eqnarray}
where $t_{ij}$ is the hopping transfer integral for fermions moving between nearest-neighbor sites,
$c_{i}^{\dagger}$ ($c_{i}$) and $f_{i}^{\dagger}$ ($f_{i}$) are,
respectively, the creation (annihilation) operators for the mobile
and trapped fermions at site $i$, and $\mu$ is the chemical potential for the mobile fermions. 
$\epsilon_i$ is the local impurity and  
$U_{i}$ denotes the local Coulomb interaction strength between trapped and mobile fermions. 
These two aforementioned quantities are randomly distributed through the lattice, being
characterized by a probability distribution function of the form  
$P (\epsilon_i) = \Theta(\Delta/2 - \epsilon_i)/\Delta$ and 
$\tilde{P}(U_i)= \Theta(\delta/2-\vert U_{i}+U \vert ) /\delta$, where 
$\Theta$ is the step function, 
$\Delta$ ($\delta$) measures the amount of Anderson (Coulomb) disorder, 
and $U$ is the mean value of the Coulomb interaction strength.  
Here, we deal only with a repulsive interaction, $U_{i} \geq 0$, which leads to $U\geq \delta/2$.  
The mean particle number for the mobile and trapped fermions at the $i$th site are given by 
$n_i =\langle c_{i}^{+}c_{i}\rangle$ and $p_i = \langle f_{i}^{+}f_{i}\rangle$, respectively, 
and are independent from each other. 

The changes caused by Coulomb disorder in the phase diagram of the 
Anderson-Falicov-Kimball model, i.e. how it drives the MIT, will be explored in the framework of 
DMFT in the following sections.

\subsection{Dynamical mean-field theory}

The equations of motion 
for Hamiltonian (\ref{eq01}) are expressed by \cite{metzner89, georges96} 
\begin{eqnarray}
&&(\omega - \epsilon_i + \mu)G_{ij}(\omega)- \sum_{l} t_{il} G_{lj}(\omega) =  \delta_{ij} + U_{i} \Gamma_{ij}(\omega), \label{eq02}\\
&&(\omega - \epsilon_i + \mu - U_{i})\Gamma_{ij}(\omega)-\sum_{l} t_{il} \Gamma_{lj}(\omega)=\delta_{ij} p_{i}, \label{eq03}  
\end{eqnarray}
where $G_{ij}(\omega)=\langle\langle c_i|c_{j}^{\dagger} \rangle\rangle_{\omega}$
and 
$\Gamma_{ij}(\omega)=\langle\langle f_{i}f_{i}^{\dagger}c_i|c_{j}^{\dagger} \rangle\rangle_{\omega}$ are the Green's functions for the single and double particle states \cite{zubarev60}, and $\delta_{ij}$ is the Kronecker delta function.
Herein we assume the lattice to be homogeneous, that is $p_{i} = p$ with $p \in [0,1]$.

According to the DMFT scheme, the eigenenergies are defined as
\begin{equation}
\Lambda (\omega )\equiv U_{i} \frac{ \Gamma_{ij}(\omega ) }{G_{ij}(\omega)}, \label{eq04}
\end{equation}
where $\Lambda(\omega)$ depends implicitly on $\epsilon_{i}$.
The hybridization function $\eta(\omega)$ is introduced through
\begin{eqnarray}
\sum_{l} t_{il} G_{lj}(\omega ) \equiv \eta (\omega) G_{ij}(\omega), \label{eq05}\\
\sum_{l} t_{il} \Gamma_{lj}(\omega ) \equiv \eta (\omega) \Gamma_{ij}(\omega). \label{eq06}
\end{eqnarray}
Now inserting Eqs. (\ref{eq04}), (\ref{eq05}), and (\ref{eq06}) into Eqs. (\ref{eq02}) and (\ref{eq03}) we obtain 
\begin{eqnarray}
G_{ij}  (\omega) &=& \frac{\delta_{ij}}{\omega-\epsilon_i + \mu - \eta(\omega) - \Lambda(\omega)}, \label{eq07}\\
\Lambda(\omega) &=& p U_i + \frac{U_i^2 (1-p)}{\omega-\epsilon_i + \mu - U_i (1-p) - \eta(\omega)}.\label{eq08}
\end{eqnarray}
According to these equations, the lattice is mapped to a set of impurity problems, each one with a random value of $\epsilon_i$,
embedded in a self-consistent field. 
Then, the LDOS is given by
\begin{equation}
\rho_i (\omega) =- \frac{1}{\pi} \mathrm{Im} [G_{ii}(\omega)],
\label{eq09}
\end{equation}
which depends on $\epsilon_i$ and $U_i$.
In order to maintain self-consistency in the disordered problem, 
the Green's functions must be solved 
by taking into account the averages for
which the translational invariance can be restored. 
The above LDOS can be evaluated by using the arithmetic or geometric mean given by, respectively,  
\begin{eqnarray}
\rho_{\mathrm{arith}} (\omega ) &=&  \int du \int d\epsilon P(\epsilon ) \tilde{P}(u) \rho (\omega, \epsilon ,u),\label{eq10} \\
\rho_{\mathrm{geom}} (\omega ) &=& \exp \left[ \int du \int d\epsilon P(\epsilon ) \tilde{P}(u) \ln \rho (\omega, \epsilon ,u) \right]. \label{eq11}
\end{eqnarray}
The translational invariant Green's function is given by the Hilbert transform
\begin{equation}
G(\omega)= \int d\omega'\frac{\rho_{m}(\omega')}{\omega-\omega'},
\label{eq12}
\end{equation}
in which $m$ denotes the type of mean being used.
The self-consistent DMFT equations are closed through
\begin{equation}
\rho_{i}(\omega) =-\frac{s}{\pi}\frac{\alpha_i^2 + s^2 + (U_i/2)^2}{[\alpha_i^2 + s^2 + (U_i/2)^2]^2-(U_i\alpha_i)^2},
\label{eq13}
\end{equation}
where $\alpha_i = \omega - \epsilon_i-r$, $r(\omega)$ and $s(\omega)$
are, respectively, the real and imaginary parts of $\eta(\omega)$ and we have
assumed that our system is defined in a
Bethe lattice. Therefore, $\eta(\omega) = G(\omega)/16$, where the energy is expressed in units of the band width.
In Eq. (\ref{eq13}), it was assumed that the band is half-filled, i.e.,  
$n_{i} = n = 1/2$ and $p=1/2$. In addition, the chemical potential was set to 
$\mu = U/2$ in order to fix the band center to lie in $\omega = 0$.

\subsection{Linearized DMFT}

The quantum states corresponding to the band center can determine 
the ground-state properties for the half-filled case. In the MIT, for example, the LDOS vanishes at this point. 
When the system is in the metallic phase, the LDOS is arbitrarily small in the vicinity of the MIT region. Hence, the transition points on the phase diagram can be determined by linearizing the DMFT equations \cite{dobro97, bulla00, souza07}.  

In the band center, the Green's function is  purely imaginary, 
$G (0) =- i \pi \rho_{m} (0)$, due to the symmetry of $\rho_{m}(\omega)$ thus leading to the recursive relation $G(0)^{(j+1)} = -i\pi \rho_{m}^{(j)}(0)$.
For $s (0) << 1$, by using Eq. (\ref{eq13}) we can obtain
the DMFT recursive relation
\begin{equation}
\rho_{m}^{(j+1)}(0)= \frac{1}{16} \rho_{m}^{(j)} (0) \Upsilon (\epsilon_i,U_i), \label{eq14}
\end{equation}
where
\begin{equation}
\Upsilon (\epsilon_i , U_i ) = \frac{\epsilon_i^2 + (U_{i}/2)^2} {( \epsilon_i^2-(U_{i}/2)^2 )^2}. \label{eq15}
\end{equation}
The boundary between metallic and insulating phases can be obtained when
the condition $\rho_{m}^{(j+1)}(0) = \rho_{m}^{(j)}(0)$ is satisfied.
Then, by using Eqs. (\ref{eq14}) and (\ref{eq15}) and evaluating the averages on $\epsilon_{i}$ and $U_{i}$, we finally obtain the expressions which determine the MIT for both arithmetic and geometric means, respectively,
\begin{equation} \label{critical_arith}
8\delta= \dfrac{1}{\Delta } \ln \left\vert \frac{U^2 -(\delta /2 -\Delta)^2}{U^2 -(\delta /2 +\Delta)^2} \right\vert,
\end{equation}
\begin{align} \label{critical_geom}
\delta \ln 4 = & \int_{U-\delta/2}^{U+\delta/2} du \left[ \ln 
\frac{u^2 + \Delta^2}{(u^2 - \Delta^2)^2} + \right. \nonumber\\
&+ \left.  \frac{2u}{\Delta} \left(\tan^{-1}\frac{\Delta}{u} - \ln \left|\frac{\Delta+u}{\Delta-u}\right|\right)\right].  
\end{align}

\section{\label{secIII}Results}

Now we investigate the system's phase diagrams obtained
from the DFMT scheme discussed in the previous section. The LDOS was evaluated for a variety of parameters as  
it gives information about the allowed states on the system. For instance, the ground-state
properties can be analysed from the following outcomes
of $\rho_{\mathrm{arith}}$ and $\rho_{\mathrm{geom}}$ at the band center $\omega = 0$ \cite{gusmao08}: i) 
$\rho_{\mathrm{arith}}(0) \neq 0$ and $\rho_{\mathrm{geom}}(0) \neq 0$ denote a metallic phase; 
ii) $\rho_{\mathrm{arith}}(0) = 0$, $\rho_{\mathrm{geom}}(0) = 0$, 
and $\int \rho_{\mathrm{geom}}(\omega)d\omega \neq 0$ indicate a Mott-insulator phase; and
iii) for $\rho_{\mathrm{arith}}(0) \neq 0$ and $\int \rho_{\mathrm{geom}}(\omega)d\omega = 0$ there is Anderson localization without a Mott gap. There are also coexistent phases which will not be highlighted in this work.
All these situations occur for appropriate sets of $U$, $\Delta$, and $\delta$ and give rise to a rich
phase diagram. 

\begin{figure*}[b!] 
\includegraphics[width=0.8\textwidth]{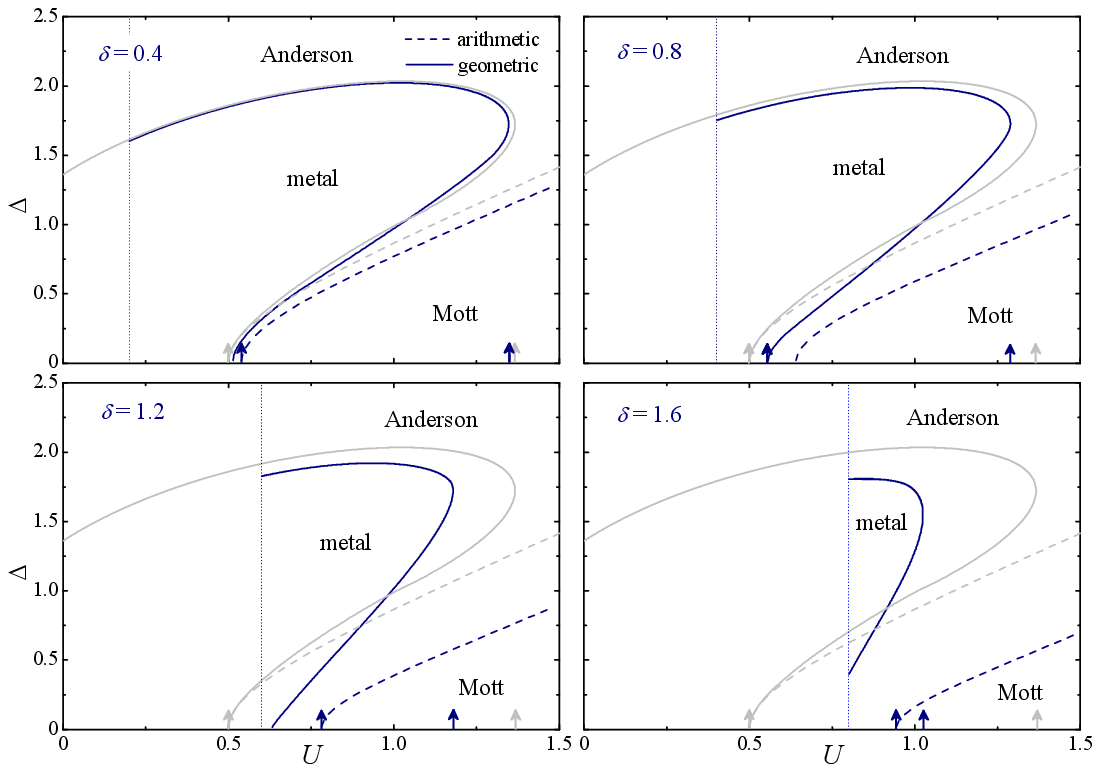}
\caption{\label{fig1} (Color online) Phase diagrams of the Anderson-Falicov-Kimball model for different Coulomb disorder strengths (dark blue lines) compared with the $\delta = 0$ case (light grey lines). 
The critical curves were obtained directly from Eqs. (\ref{critical_arith}) and (\ref{critical_geom}), regarding 
the arithmetical and geometrical means, respectively. 
The vertical dotted line splits the regions for which $U < \delta/2$ 
(left side, not considered) and $U \geq \delta/2$ (right side).
The arrows located at the $U$-axis indicate the Coulomb interaction regimes.
}
\end{figure*}

In the case where no Coulomb disorder is considered ($\delta = 0$) \cite{byczuk05, gusmao08}, 
the metallic phase is identified for small values of $U$ and $\Delta$, the Mott-insulator phase 
stabilizes as we increase $U$, and  
Anderson localization naturally overcomes for large $\Delta$. 
This is clearly seen in Fig. \ref{fig1} where we compare the phase diagrams evaluated at the band center 
with and without Coulomb disorder.
%
For the latter case, three different interaction regimes can be identified \cite{byczuk05} regarding the metallic and Mott phase boundaries (see the small arrows at the $U$-axis in Fig. \ref{fig1}): weak ($0<U<0.5$), 
intermediate ($0.5<U \lesssim 1.36$), and strong ($U \gtrsim 1.36$). 
When $\delta \neq 0$, these regimes still hold but for different critical values, with the intermediate one shrinking as $\delta$ increases. The left side of the vertical dotted lines indicates a non-physical region as we are dealing with $U \geq \delta/2$ only. 
Thus, in Fig. \ref{fig1} we see the major effect of including Coulomb disorder into the problem, which is to drive the system
to the Anderson localized phase. 

All the critical curves presented in Fig. \ref{fig1} 
were obtained directly from Eqs. (\ref{critical_arith}) and (\ref{critical_geom}),
however, it is worth pointing out that the numerical results obtained by solving the self-consistent equations 
of DMFT absolutely matches with those from the linearised DMFT. 

The above analysis was performed for the band center. Nevertheless, it is relevant to also 
observe the effects of Coulomb disorder
in the whole band. For this, we study the LDOS by proceeding with a similar 
analysis for the entire set
of ​​$\omega$ solving the self-consistent equations of DMFT (without resorting to the linearized DMFT).

\begin{figure}[t] 
\includegraphics[width=0.45\textwidth]{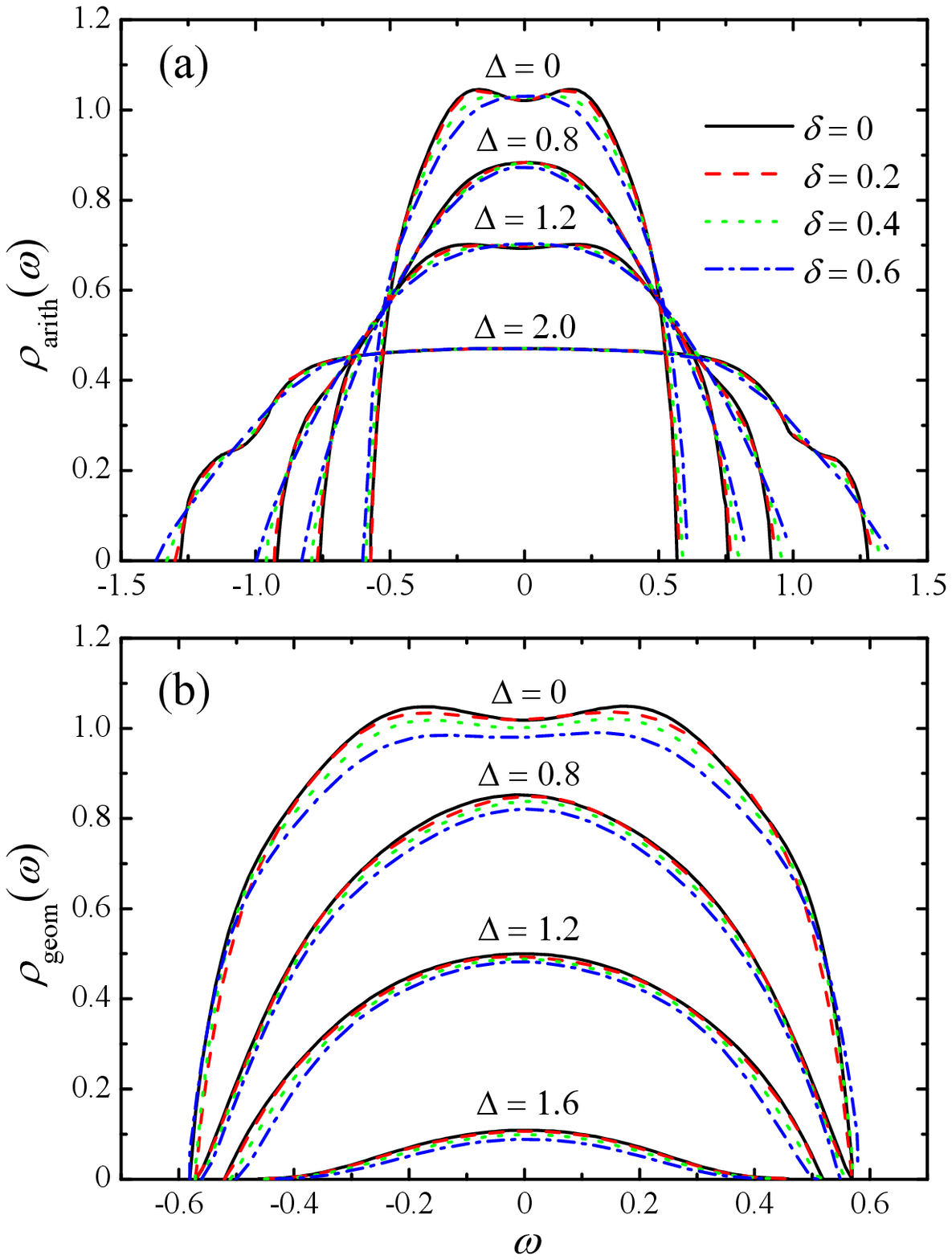}
\caption{\label{fig2} (Color online) Local density of states (LDOS) at the weak interaction regime (we set $U = 0.3$) for a range of Anderson and Coulomb disorder parameters, $\Delta$ and $\delta$, respectively, regarding the (a) arithmetic and (b) geometric mean.
Note that $\rho_{\mathrm{arith}}$ is always normalized while $\rho_{\mathrm{geom}}$ is not.    
}
\end{figure}

At the weak interaction regime, the Mott gap is not available and the system is in the metallic phase as shown in Fig. \ref{fig2}. 
By increasing $\Delta$ we reduce
the density of states at the center of the band, $\omega=0$, thus making the spectrum of gapless extended states narrower and expanding the total bandwidth \cite{byczuk05}.
For this weak regime, Coulomb disorder has practically no influence on the density of states. Even for 
large $\delta$,
the bandwidth slightly increases (decreases) for the arithmetic (geometric) case.

%

\begin{figure}[b] 
\includegraphics[width=0.45\textwidth]{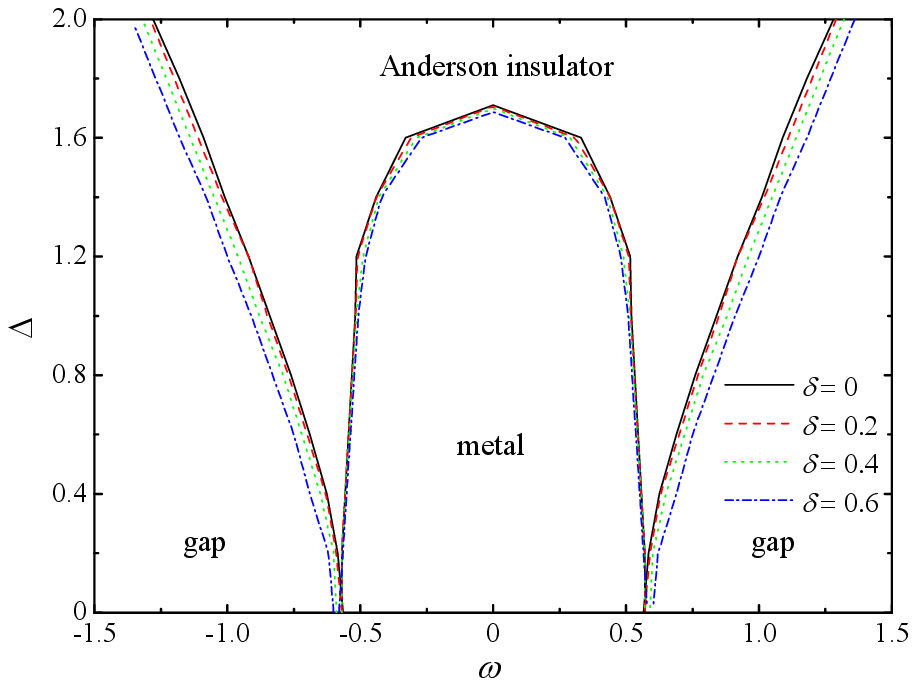}
\caption{\label{fig3} (Color online) Spectral phase diagram for the weak interaction regime ($U = 0.3$) 
and different Coulomb disorder strengths. The boundaries delimiting the gap (metallic) region was obtained using the arithmetic (geometric) mean.      
}
\end{figure}

For fixed $\Delta$ and $\delta$ values, it is possible to obtain the bandwidth by assigning the frequencies $\omega$ at which $\rho_{m}(\omega)$ vanishes.
By performing this process for several disorder strengths one can obtain a spectral phase diagram 
like the one shown in Fig. \ref{fig3}.
Localized states with no gap (Anderson insulator) can be detected in a thin energy band between the band gap and metallic states for $\Delta = 0$ and $\delta \neq 0$.
However, this phase diagram is qualitatively similar to that obtained for the Anderson model with no interaction \cite{dobro03, byczuk05}.
Therefore, at the weak interaction regime, a uniform or disordered distribution of Coulomb coupling strengths have 
no major effects in the system's properties.

\begin{figure}[t!] 
\includegraphics[width=0.45\textwidth]{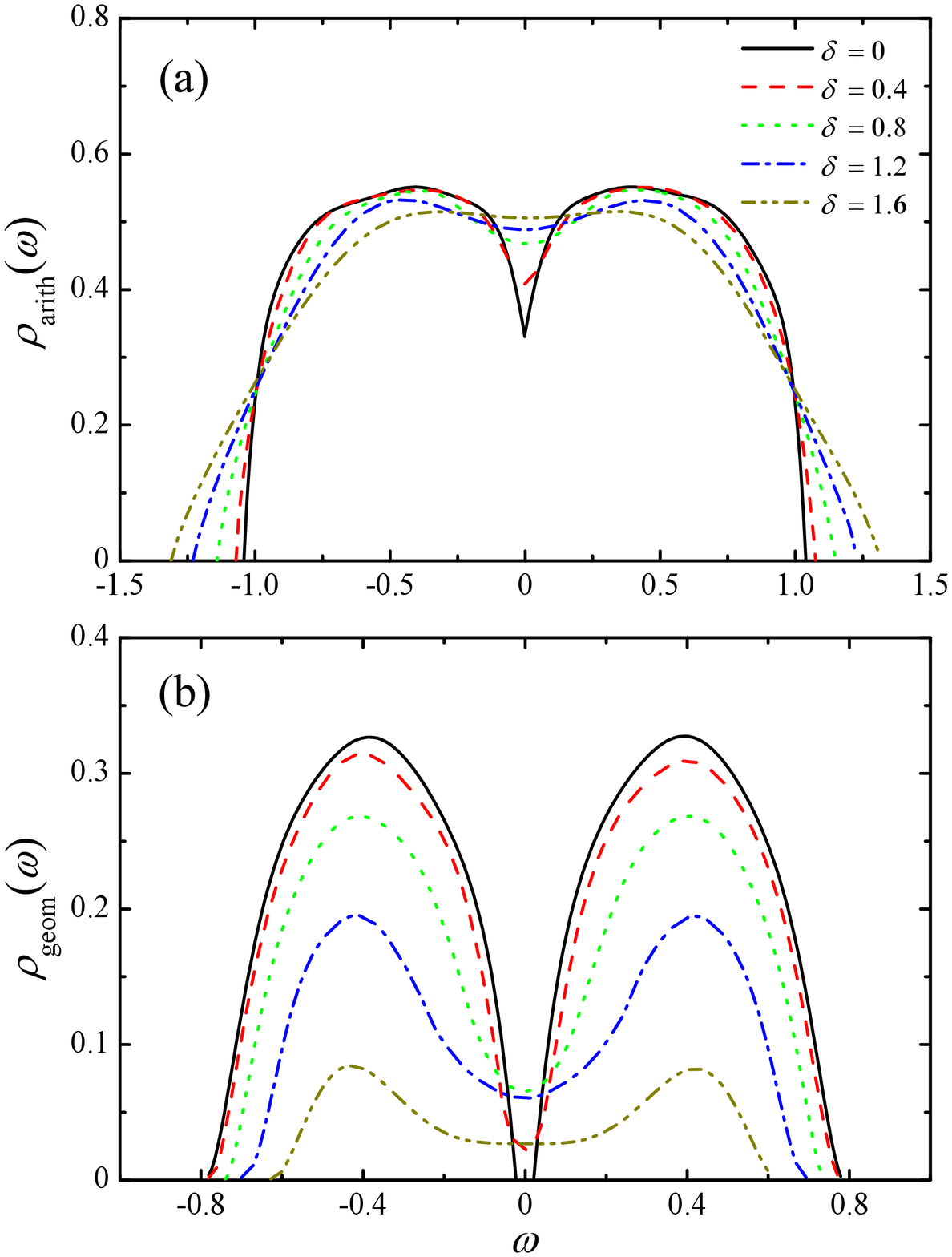}
\caption{\label{fig4} (Color online) Same as Fig. \ref{fig2} but for $U=0.9$ (intermediate interaction regime) and fixed $\Delta = 0.8$.
}
\end{figure}

As $U$ increases, the Mott gap starts to rise. At this point, we have reached the intermediate interaction regime  
as depicted earlier in Fig. \ref{fig1}. Both sources of disorder can take the system out of this Mott phase. In contrast with 
the previous weak interaction case, Coulomb disorder now has a significant role on the spectral density as seen in Fig. \ref{fig4}
for a fixed amount of Anderson disorder $\Delta$. Coulomb disorder decreases the LDOS [this effect turns to be clearer
for the geometric mean in Fig. \ref{fig4}(b)]. In addition, it makes the band length larger (smaller) when considering
the arithmetic (geometric) mean.
%
%

%
\begin{figure}[b!] 
\includegraphics[width=0.45\textwidth]{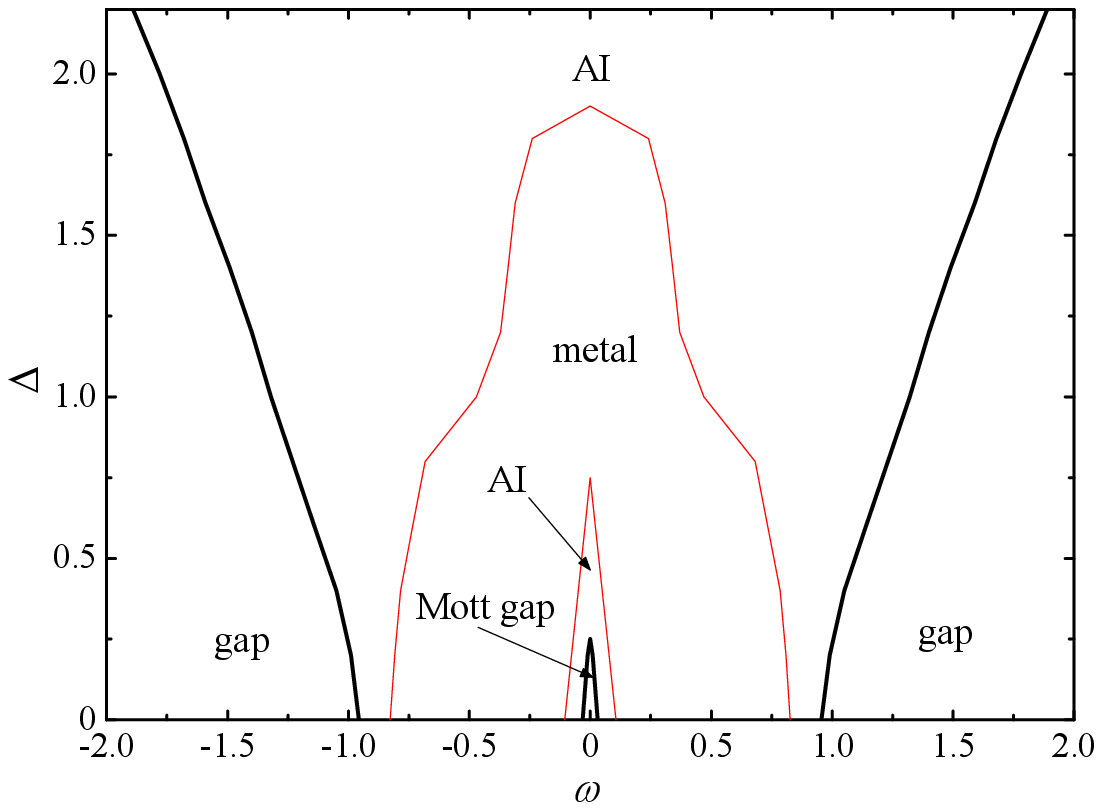}
\caption{\label{fig5} (Color online) Spectral phase diagram for the intermediate interaction regime ($U=0.9$) and $\delta = 1.2$. Thick black (thin red) curves were obtained using the arithmetic (geometric) mean and AI stands for Anderson insulator. 
}
\end{figure}
In Fig. \ref{fig5}, we show the spectral phase diagram for this regime. As expected, we now identify a Mott gap around the band center inside the Anderson insulator region. For increasing $\delta$, there is a natural tendency for localization as the Mott gap vanishes. Also, the connection between the band gap and extended (metallic) states is suppressed, thus allowing localized states in between. 
    
\begin{figure}[t!] 
\includegraphics[width=0.45\textwidth]{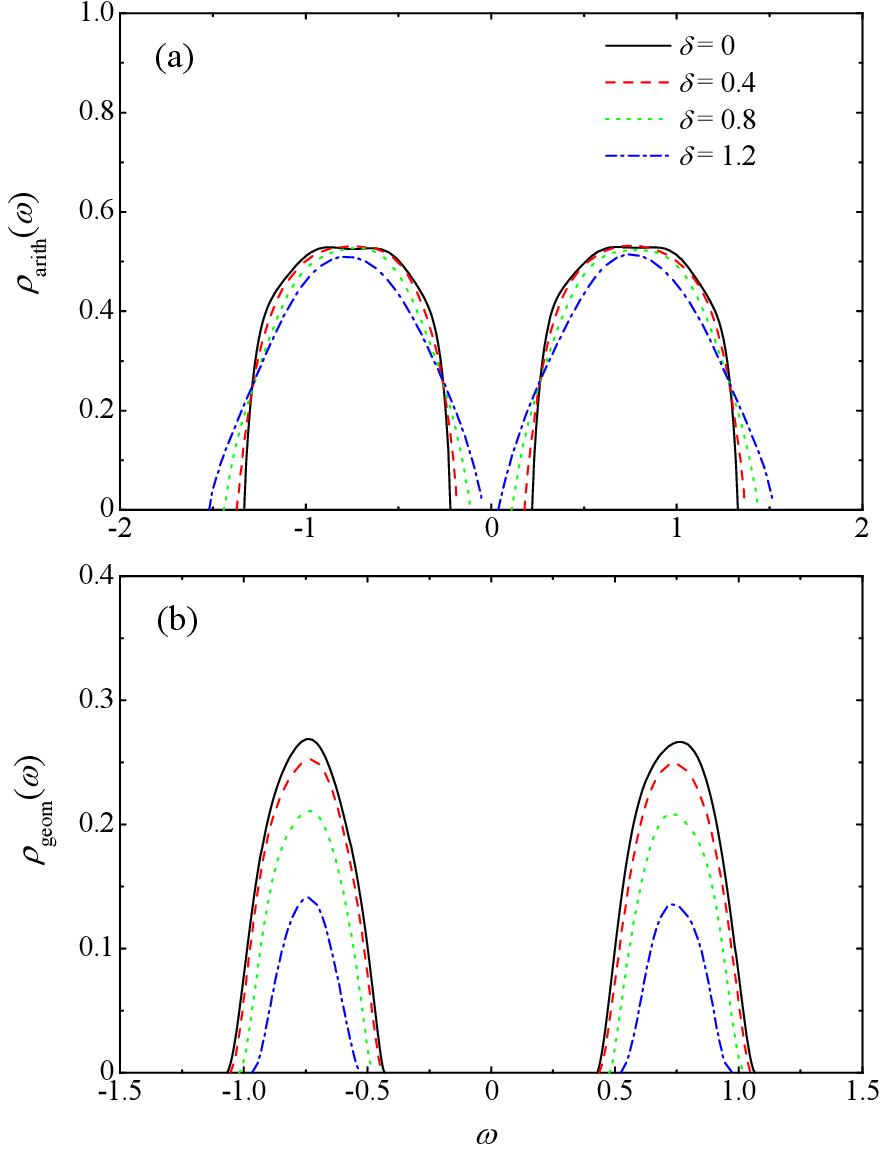}
\caption{\label{fig6} (Color online) Same as Fig. \ref{fig2} but for $U=1.5$ (strong interaction regime) and fixed $\Delta = 0.8$.
}
\end{figure}

Finally, for the strong interaction regime, Fig. \ref{fig6} shows that there are two separate bands that tend to merge with each other as $\delta$ increases and we keep $\Delta$ fixed, when considering the arithmetic mean [Fig. \ref{fig6}(a)]. For higher amounts of $\Delta$, this effect would naturally occur for 
a weaker Coulomb disorder. For the LDOS evaluated by the geometric mean [Fig. \ref{fig6}(b)], the Mott gap is not filled regardless of the disorder (of any kind) intensity. There are two branches apart which correspond to the higher and lower Hubbard sub-bands \cite{byczuk05}.

%

\begin{figure}[b!] 
\includegraphics[width=0.45\textwidth]{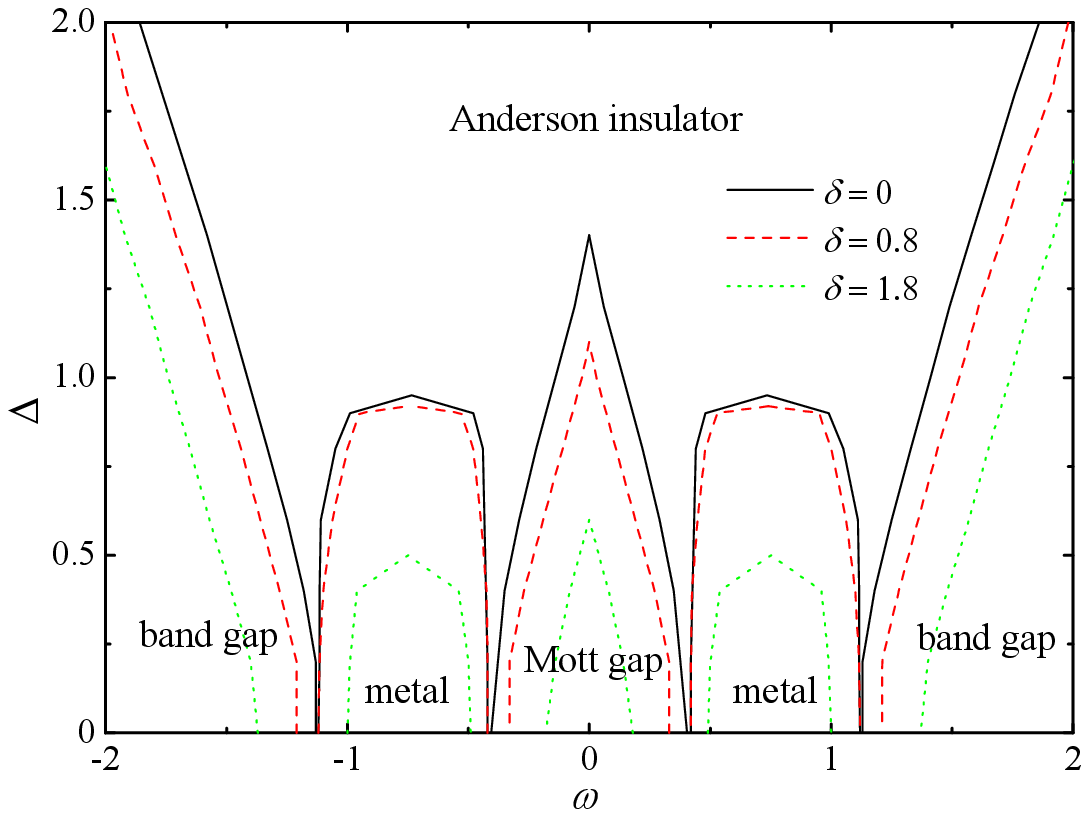}
\caption{\label{fig7} (Color online) Spectral phase diagram for the strong interaction regime ($U=1.5$) and several Couloumb
 disorder strengths $\delta$. The boundaries delimiting the Mott and band gaps were obtained using the arithmetic mean while
 the metallic region was evaluated using the geometric mean.}
\end{figure}

Figure \ref{fig7} shows the spectral phase diagram where the extended states lie in two separate lobes. 
These are surrounded by gapless localized states (Anderson insulator) when Coulomb disorder is taken into account. 
Once again we note that $\delta$ drives the system to the localized phase as both Mott gap and metallic regions shrink. 

%

\section{\label{secIV}Conclusions}

Phase diagrams describing
the metal-insulator phase transition in the Falicov-Kimball model with Anderson and Coulomb disorder 
were obtained by means of the DMFT. 
Coulomb disorder has been shown to add new features into the problem
as it directly affects the spectral density by decreasing its intensity 
and increasing the
total bandwidth as $\delta$ is increased.
Even in the absence of structural disorder, it is possible to observe localized states with no gap.
We showed that the critical values associated with the MIT 
change as the intensity of Coulomb disorder is tuned, thus turning
the system insulator-like
if this source of disorder and/or the electronic interaction are strong enough.
Therefore, in realistic situations, Coulomb disorder cannot be discarded as it 
has a significant role in the metal-insulator phase transition.

\section*{Acknowledgments}
This work was supported by CNPq (Brazilian agency).


\begin{thebibliography}{28}%
\makeatletter
\providecommand \@ifxundefined [1]{%
 \@ifx{#1\undefined}
}%
\providecommand \@ifnum [1]{%
 \ifnum #1\expandafter \@firstoftwo
 \else \expandafter \@secondoftwo
 \fi
}%
\providecommand \@ifx [1]{%
 \ifx #1\expandafter \@firstoftwo
 \else \expandafter \@secondoftwo
 \fi
}%
\providecommand \natexlab [1]{#1}%
\providecommand \enquote  [1]{``#1''}%
\providecommand \bibnamefont  [1]{#1}%
\providecommand \bibfnamefont [1]{#1}%
\providecommand \citenamefont [1]{#1}%
\providecommand \href@noop [0]{\@secondoftwo}%
\providecommand \href [0]{\begingroup \@sanitize@url \@href}%
\providecommand \@href[1]{\@@startlink{#1}\@@href}%
\providecommand \@@href[1]{\endgroup#1\@@endlink}%
\providecommand \@sanitize@url [0]{\catcode `\\12\catcode `\$12\catcode
  `\&12\catcode `\#12\catcode `\^12\catcode `\_12\catcode `\%12\relax}%
\providecommand \@@startlink[1]{}%
\providecommand \@@endlink[0]{}%
\providecommand \url  [0]{\begingroup\@sanitize@url \@url }%
\providecommand \@url [1]{\endgroup\@href {#1}{\urlprefix }}%
\providecommand \urlprefix  [0]{URL }%
\providecommand \Eprint [0]{\href }%
\providecommand \doibase [0]{http://dx.doi.org/}%
\providecommand \selectlanguage [0]{\@gobble}%
\providecommand \bibinfo  [0]{\@secondoftwo}%
\providecommand \bibfield  [0]{\@secondoftwo}%
\providecommand \translation [1]{[#1]}%
\providecommand \BibitemOpen [0]{}%
\providecommand \bibitemStop [0]{}%
\providecommand \bibitemNoStop [0]{.\EOS\space}%
\providecommand \EOS [0]{\spacefactor3000\relax}%
\providecommand \BibitemShut  [1]{\csname bibitem#1\endcsname}%
\let\auto@bib@innerbib\@empty
\bibitem [{\citenamefont {Imada}\ \emph {et~al.}(1998)\citenamefont {Imada},
  \citenamefont {Fujimori},\ and\ \citenamefont {Tokura}}]{imada98}%
  \BibitemOpen
  \bibfield  {author} {\bibinfo {author} {\bibfnamefont {M.}~\bibnamefont
  {Imada}}, \bibinfo {author} {\bibfnamefont {A.}~\bibnamefont {Fujimori}}, \
  and\ \bibinfo {author} {\bibfnamefont {Y.}~\bibnamefont {Tokura}},\ }\href
  {\doibase 10.1103/RevModPhys.70.1039} {\bibfield  {journal} {\bibinfo
  {journal} {Rev. Mod. Phys.}\ }\textbf {\bibinfo {volume} {70}},\ \bibinfo
  {pages} {1039} (\bibinfo {year} {1998})}\BibitemShut {NoStop}%
\bibitem [{\citenamefont {Anderson}(1961)}]{anderson61}%
  \BibitemOpen
  \bibfield  {author} {\bibinfo {author} {\bibfnamefont {P.~W.}\ \bibnamefont
  {Anderson}},\ }\href {\doibase 10.1103/PhysRev.124.41} {\bibfield  {journal}
  {\bibinfo  {journal} {Phys. Rev.}\ }\textbf {\bibinfo {volume} {124}},\
  \bibinfo {pages} {41} (\bibinfo {year} {1961})}\BibitemShut {NoStop}%
\bibitem [{\citenamefont {Mott}(1949)}]{mott49}%
  \BibitemOpen
  \bibfield  {author} {\bibinfo {author} {\bibfnamefont {N.~F.}\ \bibnamefont
  {Mott}},\ }\href@noop {} {\bibfield  {journal} {\bibinfo  {journal} {Proc.
  Phys. Soc., London, Sect. A}\ }\textbf {\bibinfo {volume} {62}},\ \bibinfo
  {pages} {416} (\bibinfo {year} {1949})}\BibitemShut {NoStop}%
\bibitem [{\citenamefont {Lee}\ and\ \citenamefont
  {Ramakrishnan}(1985)}]{lee85}%
  \BibitemOpen
  \bibfield  {author} {\bibinfo {author} {\bibfnamefont {P.~A.}\ \bibnamefont
  {Lee}}\ and\ \bibinfo {author} {\bibfnamefont {T.~V.}\ \bibnamefont
  {Ramakrishnan}},\ }\href {\doibase 10.1103/RevModPhys.57.287} {\bibfield
  {journal} {\bibinfo  {journal} {Rev. Mod. Phys.}\ }\textbf {\bibinfo {volume}
  {57}},\ \bibinfo {pages} {287} (\bibinfo {year} {1985})}\BibitemShut
  {NoStop}%
\bibitem [{\citenamefont {Belitz}\ and\ \citenamefont
  {Kirkpatrick}(1994)}]{belitz94}%
  \BibitemOpen
  \bibfield  {author} {\bibinfo {author} {\bibfnamefont {D.}~\bibnamefont
  {Belitz}}\ and\ \bibinfo {author} {\bibfnamefont {T.~R.}\ \bibnamefont
  {Kirkpatrick}},\ }\href {\doibase 10.1103/RevModPhys.66.261} {\bibfield
  {journal} {\bibinfo  {journal} {Rev. Mod. Phys.}\ }\textbf {\bibinfo {volume}
  {66}},\ \bibinfo {pages} {261} (\bibinfo {year} {1994})}\BibitemShut
  {NoStop}%
\bibitem [{\citenamefont {Byczuk}(2005)}]{byczuk05}%
  \BibitemOpen
  \bibfield  {author} {\bibinfo {author} {\bibfnamefont {K.}~\bibnamefont
  {Byczuk}},\ }\href {\doibase 10.1103/PhysRevB.71.205105} {\bibfield
  {journal} {\bibinfo  {journal} {Phys. Rev. B}\ }\textbf {\bibinfo {volume}
  {71}},\ \bibinfo {pages} {205105} (\bibinfo {year} {2005})}\BibitemShut
  {NoStop}%
\bibitem [{\citenamefont {Byczuk}\ \emph {et~al.}(2005)\citenamefont {Byczuk},
  \citenamefont {Hofstetter},\ and\ \citenamefont {Vollhardt}}]{byczuk05-2}%
  \BibitemOpen
  \bibfield  {author} {\bibinfo {author} {\bibfnamefont {K.}~\bibnamefont
  {Byczuk}}, \bibinfo {author} {\bibfnamefont {W.}~\bibnamefont {Hofstetter}},
  \ and\ \bibinfo {author} {\bibfnamefont {D.}~\bibnamefont {Vollhardt}},\
  }\href {\doibase 10.1103/PhysRevLett.94.056404} {\bibfield  {journal}
  {\bibinfo  {journal} {Phys. Rev. Lett.}\ }\textbf {\bibinfo {volume} {94}},\
  \bibinfo {pages} {056404} (\bibinfo {year} {2005})}\BibitemShut {NoStop}%
\bibitem [{\citenamefont {Souza}\ \emph {et~al.}(2007)\citenamefont {Souza},
  \citenamefont {Maionchi},\ and\ \citenamefont {Herrmann}}]{souza07}%
  \BibitemOpen
  \bibfield  {author} {\bibinfo {author} {\bibfnamefont {A.~M.~C.}\
  \bibnamefont {Souza}}, \bibinfo {author} {\bibfnamefont {D.~O.}\ \bibnamefont
  {Maionchi}}, \ and\ \bibinfo {author} {\bibfnamefont {H.~J.}\ \bibnamefont
  {Herrmann}},\ }\href {\doibase 10.1103/PhysRevB.76.035111} {\bibfield
  {journal} {\bibinfo  {journal} {Phys. Rev. B}\ }\textbf {\bibinfo {volume}
  {76}},\ \bibinfo {pages} {035111} (\bibinfo {year} {2007})}\BibitemShut
  {NoStop}%
\bibitem [{\citenamefont {Gusm\~ao}(2008)}]{gusmao08}%
  \BibitemOpen
  \bibfield  {author} {\bibinfo {author} {\bibfnamefont {M.~A.}\ \bibnamefont
  {Gusm\~ao}},\ }\href {\doibase 10.1103/PhysRevB.77.245116} {\bibfield
  {journal} {\bibinfo  {journal} {Phys. Rev. B}\ }\textbf {\bibinfo {volume}
  {77}},\ \bibinfo {pages} {245116} (\bibinfo {year} {2008})}\BibitemShut
  {NoStop}%
\bibitem [{\citenamefont {Maionchi}\ \emph {et~al.}(2008)\citenamefont
  {Maionchi}, \citenamefont {Souza}, \citenamefont {Herrmann},\ and\
  \citenamefont {da~Costa~Filho}}]{maionchi08}%
  \BibitemOpen
  \bibfield  {author} {\bibinfo {author} {\bibfnamefont {D.~O.}\ \bibnamefont
  {Maionchi}}, \bibinfo {author} {\bibfnamefont {A.~M.~C.}\ \bibnamefont
  {Souza}}, \bibinfo {author} {\bibfnamefont {H.~J.}\ \bibnamefont {Herrmann}},
  \ and\ \bibinfo {author} {\bibfnamefont {R.~N.}\ \bibnamefont
  {da~Costa~Filho}},\ }\href {\doibase 10.1103/PhysRevB.77.245126} {\bibfield
  {journal} {\bibinfo  {journal} {Phys. Rev. B}\ }\textbf {\bibinfo {volume}
  {77}},\ \bibinfo {pages} {245126} (\bibinfo {year} {2008})}\BibitemShut
  {NoStop}%
\bibitem [{\citenamefont {Orignac}\ \emph {et~al.}(2006)\citenamefont
  {Orignac}, \citenamefont {Rosso}, \citenamefont {Chitra},\ and\ \citenamefont
  {Giamarchi}}]{orignac06}%
  \BibitemOpen
  \bibfield  {author} {\bibinfo {author} {\bibfnamefont {E.}~\bibnamefont
  {Orignac}}, \bibinfo {author} {\bibfnamefont {A.}~\bibnamefont {Rosso}},
  \bibinfo {author} {\bibfnamefont {R.}~\bibnamefont {Chitra}}, \ and\ \bibinfo
  {author} {\bibfnamefont {T.}~\bibnamefont {Giamarchi}},\ }\href {\doibase
  10.1103/PhysRevB.73.035112} {\bibfield  {journal} {\bibinfo  {journal} {Phys.
  Rev. B}\ }\textbf {\bibinfo {volume} {73}},\ \bibinfo {pages} {035112}
  (\bibinfo {year} {2006})}\BibitemShut {NoStop}%
\bibitem [{\citenamefont {Shklovskii}(2007)}]{shklovskii07}%
  \BibitemOpen
  \bibfield  {author} {\bibinfo {author} {\bibfnamefont {B.~I.}\ \bibnamefont
  {Shklovskii}},\ }\href {\doibase 10.1103/PhysRevB.76.224511} {\bibfield
  {journal} {\bibinfo  {journal} {Phys. Rev. B}\ }\textbf {\bibinfo {volume}
  {76}},\ \bibinfo {pages} {224511} (\bibinfo {year} {2007})}\BibitemShut
  {NoStop}%
\bibitem [{\citenamefont {Hwang}\ and\ \citenamefont
  {Das~Sarma}(2009)}]{hwang09}%
  \BibitemOpen
  \bibfield  {author} {\bibinfo {author} {\bibfnamefont {E.~H.}\ \bibnamefont
  {Hwang}}\ and\ \bibinfo {author} {\bibfnamefont {S.}~\bibnamefont
  {Das~Sarma}},\ }\href {\doibase 10.1103/PhysRevB.79.165404} {\bibfield
  {journal} {\bibinfo  {journal} {Phys. Rev. B}\ }\textbf {\bibinfo {volume}
  {79}},\ \bibinfo {pages} {165404} (\bibinfo {year} {2009})}\BibitemShut
  {NoStop}%
\bibitem [{\citenamefont {Das~Sarma}\ and\ \citenamefont
  {Hwang}(2013)}]{sarma13}%
  \BibitemOpen
  \bibfield  {author} {\bibinfo {author} {\bibfnamefont {S.}~\bibnamefont
  {Das~Sarma}}\ and\ \bibinfo {author} {\bibfnamefont {E.~H.}\ \bibnamefont
  {Hwang}},\ }\href {\doibase 10.1103/PhysRevB.88.035439} {\bibfield  {journal}
  {\bibinfo  {journal} {Phys. Rev. B}\ }\textbf {\bibinfo {volume} {88}},\
  \bibinfo {pages} {035439} (\bibinfo {year} {2013})}\BibitemShut {NoStop}%
\bibitem [{\citenamefont {Das~Sarma}\ \emph {et~al.}(2013)\citenamefont
  {Das~Sarma}, \citenamefont {Hwang},\ and\ \citenamefont {Li}}]{sarma13-2}%
  \BibitemOpen
  \bibfield  {author} {\bibinfo {author} {\bibfnamefont {S.}~\bibnamefont
  {Das~Sarma}}, \bibinfo {author} {\bibfnamefont {E.~H.}\ \bibnamefont
  {Hwang}}, \ and\ \bibinfo {author} {\bibfnamefont {Q.}~\bibnamefont {Li}},\
  }\href {\doibase 10.1103/PhysRevB.88.155310} {\bibfield  {journal} {\bibinfo
  {journal} {Phys. Rev. B}\ }\textbf {\bibinfo {volume} {88}},\ \bibinfo
  {pages} {155310} (\bibinfo {year} {2013})}\BibitemShut {NoStop}%
\bibitem [{\citenamefont {Hubbard}(1963)}]{hubbard63}%
  \BibitemOpen
  \bibfield  {author} {\bibinfo {author} {\bibfnamefont {J.}~\bibnamefont
  {Hubbard}},\ }\href {\doibase 10.1098/rspa.1963.0204} {\bibfield  {journal}
  {\bibinfo  {journal} {Proceedings of the Royal Society of London. Series A.
  Mathematical and Physical Sciences}\ }\textbf {\bibinfo {volume} {276}},\
  \bibinfo {pages} {238} (\bibinfo {year} {1963})}\BibitemShut {NoStop}%
\bibitem [{\citenamefont {Falicov}\ and\ \citenamefont
  {Kimball}(1969)}]{falicov69}%
  \BibitemOpen
  \bibfield  {author} {\bibinfo {author} {\bibfnamefont {L.~M.}\ \bibnamefont
  {Falicov}}\ and\ \bibinfo {author} {\bibfnamefont {J.~C.}\ \bibnamefont
  {Kimball}},\ }\href {\doibase 10.1103/PhysRevLett.22.997} {\bibfield
  {journal} {\bibinfo  {journal} {Phys. Rev. Lett.}\ }\textbf {\bibinfo
  {volume} {22}},\ \bibinfo {pages} {997} (\bibinfo {year} {1969})}\BibitemShut
  {NoStop}%
\bibitem [{\citenamefont {Kennedy}\ and\ \citenamefont
  {Lieb}(1986)}]{kennedy86}%
  \BibitemOpen
  \bibfield  {author} {\bibinfo {author} {\bibfnamefont {T.}~\bibnamefont
  {Kennedy}}\ and\ \bibinfo {author} {\bibfnamefont {E.}~\bibnamefont {Lieb}},\
  }\href@noop {} {\bibfield  {journal} {\bibinfo  {journal} {Physica A}\
  }\textbf {\bibinfo {volume} {138}},\ \bibinfo {pages} {320} (\bibinfo {year}
  {1986})}\BibitemShut {NoStop}%
\bibitem [{\citenamefont {Freericks}(1993)}]{freericks93}%
  \BibitemOpen
  \bibfield  {author} {\bibinfo {author} {\bibfnamefont {J.~K.}\ \bibnamefont
  {Freericks}},\ }\href {\doibase 10.1103/PhysRevB.47.9263} {\bibfield
  {journal} {\bibinfo  {journal} {Phys. Rev. B}\ }\textbf {\bibinfo {volume}
  {47}},\ \bibinfo {pages} {9263} (\bibinfo {year} {1993})}\BibitemShut
  {NoStop}%
\bibitem [{\citenamefont {Freericks}\ and\ \citenamefont
  {Miller}(2000)}]{freericks00}%
  \BibitemOpen
  \bibfield  {author} {\bibinfo {author} {\bibfnamefont {J.~K.}\ \bibnamefont
  {Freericks}}\ and\ \bibinfo {author} {\bibfnamefont {P.}~\bibnamefont
  {Miller}},\ }\href {\doibase 10.1103/PhysRevB.62.10022} {\bibfield  {journal}
  {\bibinfo  {journal} {Phys. Rev. B}\ }\textbf {\bibinfo {volume} {62}},\
  \bibinfo {pages} {10022} (\bibinfo {year} {2000})}\BibitemShut {NoStop}%
\bibitem [{\citenamefont {Mac\^edo}\ \emph {et~al.}(2001)\citenamefont
  {Mac\^edo}, \citenamefont {Azevedo},\ and\ \citenamefont
  {de~Souza}}]{macedo01}%
  \BibitemOpen
  \bibfield  {author} {\bibinfo {author} {\bibfnamefont {C.~A.}\ \bibnamefont
  {Mac\^edo}}, \bibinfo {author} {\bibfnamefont {L.~G.}\ \bibnamefont
  {Azevedo}}, \ and\ \bibinfo {author} {\bibfnamefont {A.~M.~C.}\ \bibnamefont
  {de~Souza}},\ }\href {\doibase 10.1103/PhysRevB.64.184441} {\bibfield
  {journal} {\bibinfo  {journal} {Phys. Rev. B}\ }\textbf {\bibinfo {volume}
  {64}},\ \bibinfo {pages} {184441} (\bibinfo {year} {2001})}\BibitemShut
  {NoStop}%
\bibitem [{\citenamefont {Metzner}\ and\ \citenamefont
  {Vollhardt}(1989)}]{metzner89}%
  \BibitemOpen
  \bibfield  {author} {\bibinfo {author} {\bibfnamefont {W.}~\bibnamefont
  {Metzner}}\ and\ \bibinfo {author} {\bibfnamefont {D.}~\bibnamefont
  {Vollhardt}},\ }\href {\doibase 10.1103/PhysRevLett.62.324} {\bibfield
  {journal} {\bibinfo  {journal} {Phys. Rev. Lett.}\ }\textbf {\bibinfo
  {volume} {62}},\ \bibinfo {pages} {324} (\bibinfo {year} {1989})}\BibitemShut
  {NoStop}%
\bibitem [{\citenamefont {Georges}\ \emph {et~al.}(1996)\citenamefont
  {Georges}, \citenamefont {Kotliar}, \citenamefont {Krauth},\ and\
  \citenamefont {Rozenberg}}]{georges96}%
  \BibitemOpen
  \bibfield  {author} {\bibinfo {author} {\bibfnamefont {A.}~\bibnamefont
  {Georges}}, \bibinfo {author} {\bibfnamefont {G.}~\bibnamefont {Kotliar}},
  \bibinfo {author} {\bibfnamefont {W.}~\bibnamefont {Krauth}}, \ and\ \bibinfo
  {author} {\bibfnamefont {M.~J.}\ \bibnamefont {Rozenberg}},\ }\href {\doibase
  10.1103/RevModPhys.68.13} {\bibfield  {journal} {\bibinfo  {journal} {Rev.
  Mod. Phys.}\ }\textbf {\bibinfo {volume} {68}},\ \bibinfo {pages} {13}
  (\bibinfo {year} {1996})}\BibitemShut {NoStop}%
\bibitem [{\citenamefont {Freericks}\ and\ \citenamefont
  {Zlati\ifmmode~\acute{c}\else \'{c}\fi{}}(2003)}]{freericks03rev}%
  \BibitemOpen
  \bibfield  {author} {\bibinfo {author} {\bibfnamefont {J.~K.}\ \bibnamefont
  {Freericks}}\ and\ \bibinfo {author} {\bibfnamefont {V.}~\bibnamefont
  {Zlati\ifmmode~\acute{c}\else \'{c}\fi{}}},\ }\href {\doibase
  10.1103/RevModPhys.75.1333} {\bibfield  {journal} {\bibinfo  {journal} {Rev.
  Mod. Phys.}\ }\textbf {\bibinfo {volume} {75}},\ \bibinfo {pages} {1333}
  (\bibinfo {year} {2003})}\BibitemShut {NoStop}%
\bibitem [{\citenamefont {Zubarev}(1960)}]{zubarev60}%
  \BibitemOpen
  \bibfield  {author} {\bibinfo {author} {\bibfnamefont {D.~N.}\ \bibnamefont
  {Zubarev}},\ }\href@noop {} {\bibfield  {journal} {\bibinfo  {journal} {Sov.
  Phys. Usp.}\ }\textbf {\bibinfo {volume} {3}},\ \bibinfo {pages} {320}
  (\bibinfo {year} {1960})}\BibitemShut {NoStop}%
\bibitem [{\citenamefont {Dobrosavljevi\ifmmode~\acute{c}\else \'{c}\fi{}}\
  and\ \citenamefont {Kotliar}(1997)}]{dobro97}%
  \BibitemOpen
  \bibfield  {author} {\bibinfo {author} {\bibfnamefont {V.}~\bibnamefont
  {Dobrosavljevi\ifmmode~\acute{c}\else \'{c}\fi{}}}\ and\ \bibinfo {author}
  {\bibfnamefont {G.}~\bibnamefont {Kotliar}},\ }\href {\doibase
  10.1103/PhysRevLett.78.3943} {\bibfield  {journal} {\bibinfo  {journal}
  {Phys. Rev. Lett.}\ }\textbf {\bibinfo {volume} {78}},\ \bibinfo {pages}
  {3943} (\bibinfo {year} {1997})}\BibitemShut {NoStop}%
\bibitem [{\citenamefont {Bulla}\ and\ \citenamefont
  {Potthoff}(2000)}]{bulla00}%
  \BibitemOpen
  \bibfield  {author} {\bibinfo {author} {\bibfnamefont {R.}~\bibnamefont
  {Bulla}}\ and\ \bibinfo {author} {\bibfnamefont {M.}~\bibnamefont
  {Potthoff}},\ }\href@noop {} {\bibfield  {journal} {\bibinfo  {journal} {Eur.
  Phys. J. B}\ }\textbf {\bibinfo {volume} {13}},\ \bibinfo {pages} {257}
  (\bibinfo {year} {2000})}\BibitemShut {NoStop}%
\bibitem [{\citenamefont {Dobrosavljevi\ifmmode~\acute{c}\else \'{c}\fi{}}\
  \emph {et~al.}(2003)\citenamefont {Dobrosavljevi\ifmmode~\acute{c}\else
  \'{c}\fi{}}, \citenamefont {Pastor},\ and\ \citenamefont
  {Nikolic}}]{dobro03}%
  \BibitemOpen
  \bibfield  {author} {\bibinfo {author} {\bibfnamefont {V.}~\bibnamefont
  {Dobrosavljevi\ifmmode~\acute{c}\else \'{c}\fi{}}}, \bibinfo {author}
  {\bibfnamefont {A.~A.}\ \bibnamefont {Pastor}}, \ and\ \bibinfo {author}
  {\bibfnamefont {B.~K.}\ \bibnamefont {Nikolic}},\ }\href@noop {} {\bibfield
  {journal} {\bibinfo  {journal} {Europhys. Lett.}\ }\textbf {\bibinfo {volume}
  {62}},\ \bibinfo {pages} {76} (\bibinfo {year} {2003})}\BibitemShut {NoStop}%
\end{thebibliography}
%

\end{document}